\documentclass[iop]{emulateapj}
\usepackage{graphicx, amsmath,natbib}
\shorttitle{Cool Gas in the Halos of Early-type Galaxies}

\providecommand\scription[2]{\scriptsize#1$\;${\scriptsize\uppercase\expandafter{\romannumeral #2}}\relax}%
\providecommand{\Lya}{\ensuremath{\mbox{Ly}\alpha}}
\providecommand{\lya}{\ensuremath{\mbox{Ly}\alpha}}

\def\cm#1{\, {\rm cm^{#1}}}
\providecommand{\HI}{\ensuremath{\mbox{\ion{H}{1}}}}
\providecommand{\OVI}{\ensuremath{\mbox{\ion{O}{6}}}}

\providecommand{\kms}{\,\ensuremath{\rm{km\,s}^{-1}}}
\providecommand{\logNHI}{\ensuremath{\mbox{log\,}N_{\scHI}}}
\providecommand{\scHI}{\ensuremath{\mbox{\scription{H}{1}}}}
\providecommand{\NHI}{\ensuremath{N_{\scHI}}}
\providecommand{\numetg}{16}

\providecommand{\cmsq}{\,\ensuremath{\mbox{cm}^{-2}}}
\providecommand{\kpc}{\,\ensuremath{\mbox{kpc}}}
\providecommand{\mA}{\,\ensuremath{\mbox{m\AA}}}

\providecommand{\K}{\,\ensuremath{\mbox{K}}}

\def\spose#1{\hbox to 0pt{#1\hss}}
\def\simlt{\mathrel{\spose{\lower 3pt\hbox{$\mathchar"218$}}
     \raise 2.0pt\hbox{$\mathchar"13C$}}}
\def\simgt{\mathrel{\spose{\lower 3pt\hbox{$\mathchar"218$}}
     \raise 2.0pt\hbox{$\mathchar"13E$}}}

\providecommand{\Msun}{\,\ensuremath{\mbox{M}_{\odot}}}
\providecommand{\etal}{\ensuremath{\mbox{et~al.}}}

\shortauthors{Thom \etal}

\begin{document}

\title{Not Dead Yet: Cool Circumgalactic Gas in the Halos of Early Type Galaxies\altaffilmark{1}}

\author{Christopher Thom\altaffilmark{2}, 
  Jason Tumlinson\altaffilmark{2}, 
  Jessica K. Werk\altaffilmark{3}, 
  J. Xavier Prochaska\altaffilmark{3}, 
  Benjamin D. Oppenheimer\altaffilmark{4}, 
  Molly S. Peeples\altaffilmark{5},
  Todd M. Tripp\altaffilmark{6},
  Neal S. Katz\altaffilmark{6},
  John M. O'Meara\altaffilmark{7},
  Amanda Brady Ford\altaffilmark{8},
  Romeel Dav{\'e}\altaffilmark{8},
  Kenneth R. Sembach\altaffilmark{2},
  David H. Weinberg\altaffilmark{9}
}

\altaffiltext{1}{Based on observations made with the NASA/ESA Hubble Space Telescope, obtained at
  the Space Telescope Science Institute, which is operated by the Association of Universities for
  Research in Astronomy, Inc., under NASA contract NAS 5-26555. These observations are associated
  with program GO11598.}
\altaffiltext{2}{Space Telescope Science Institute, 3700 San Martin Drive, Baltimore, MD, 21218,
  USA}
\altaffiltext{3}{UCO/Lick Observatory, University of California, Santa Cruz, CA}
\altaffiltext{4}{Leiden Observatory, Leiden University, the Netherlands}
\altaffiltext{5}{Center for Galaxy Evolution Fellow, University of California Los Angeles, Los Angeles, CA}
\altaffiltext{6}{Department of Astronomy, University of Massachusetts, Amherst, MA}
\altaffiltext{7}{Department of Chemistry and Physics, Saint Michael's College, Colchester, VT}
\altaffiltext{8}{Steward Observatory, University of Arizona, Tucson, AZ}
\altaffiltext{9}{Department of Astronomy, The Ohio State University, Columbus, OH}

\begin{abstract}
  We report new observations of circumgalactic gas in the halos of early type galaxies obtained by
  the COS-Halos Survey with the Cosmic Origins Spectrograph onboard the Hubble Space Telescope.  We
  find that detections of \HI\ surrounding early type galaxies are typically as common and strong as
  around star-forming galaxies, implying that the total mass of circumgalactic material is
  comparable in the two populations. For early type galaxies, the covering fraction for
  \HI\ absorption above $10^{16}\cmsq$ is $\sim 40-50\%$ within $\sim150$ kpc. Line widths and
  kinematics of the detected material show it to be cold ($T \lesssim 10^5$ K) in comparison to the
  virial temperature of the host halos.  The implied masses of cool, photoionized CGM baryons may be
  up to $10^{9}-10^{11}\Msun$.  Contrary to some theoretical expectations, strong halo
  \HI\ absorbers do not disappear as part of the quenching of star-formation. Even passive galaxies
  retain significant reservoirs of halo baryons which could replenish the interstellar gas reservoir
  and eventually form stars.  This halo gas may feed the diffuse and molecular gas that is
  frequently observed inside ETGs.
  \end{abstract}


\section{Gas and Quenching in Early Type Galaxies}

The dichotomy of color, star-formation rate, morphology, etc.\ between spirals and early-type
galaxies (ETGs) was established nearly a century ago, and has been highly refined over the past
decade \citep[e.g.][]{baldry04,bell04,faber07}.  Despite long-standing empirical measurements, the
physical processes that divide the two populations are still uncertain. The main challenge is to
identify the mechanism(s) that quench star-formation (SF) and that inhibited it over the past $\sim
10$\,Gyr. Dramatically different scenarios have been proposed for how star formation is quenched.
Ejective feedback is generally caused by the galaxy itself, and includes super winds that expel gas
\citep[e.g.][]{sdh05} or more general heating and stripping of the gas \citep{tb09}, as well as
mergers. Conversely, preventative feedback, such as shock-heating of the gas prior to its accretion
\citep[e.g.][]{db06} or the disruption of in-falling clouds and their diffusion into the galaxy's
gaseous halo \citep[][]{putman-etal-11-head-tail-CHVCs, bland-hawthorn-etal-07-MS-Halpha}, are
consequences of ETGs living in massive halos.

\begin{figure*}[!t]
\begin{center} 
\epsscale{1.2}
\plotone{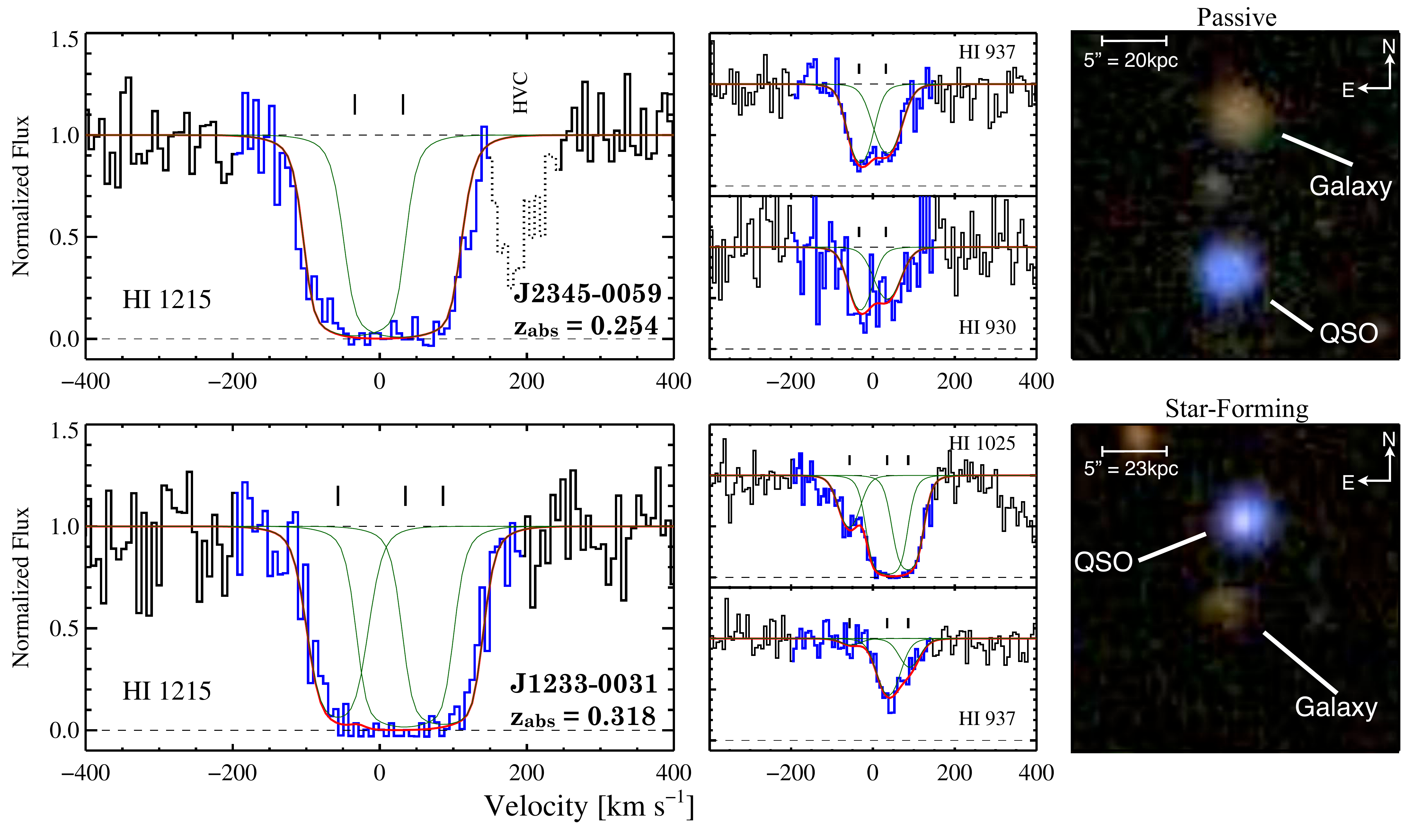}
\label{fig1}
\end{center} 
\caption{\HI\ absorption profiles and SDSS images for two example passive (top) and star-forming
  (bottom) galaxies. The strong \Lya\ absorption profiles consist of multiple blended components,
  and are similar for both galaxies. Individual Voigt profile components are shown in green;
  composite profiles are in red. The weaker Lyman series lines in the middle panels constrain the
  component structure and intrinsic line widths, which in turn constrain the gas temperature
  (\S~\ref{cold}). }
\end{figure*}

The paradigm of `red and dead' galaxies as quiescent, passively evolving stellar systems is being
reconsidered in light of findings that a non-negligible fraction of them contain gas and dust, and a
subset have experienced recent SF \citep[e.g.][]{lees91,
  gonzalez93,tfw+00,thomas10,kaviraj-10-ETG-SF}.  There is also a high incidence of field ETGs with
molecular gas \citep[22\%,][]{young11}, \HI\ gas \citep[40\%,][]{serra11}, and ionized gas
\citep[73\%,][]{davis11}. The gas masses are often comparable to those of spirals in both atomic and
molecular phases, with $M_{\rm HI} \sim 1 - 50 \times 10^7$ M$_{\odot}$ \citep{oosterloo10} and
$M_{\rm H2} = 10^{7-9}$ M$_{\odot}$ \cite[Figure~6 of][]{young11}. The corresponding SFRs reach up
to a few $M_\odot \; \rm yr^{-1}$ \citep{crocker11}. Nevertheless, most ETGs are quenched, and
remain so. These observations raise a new series of challenges: What is the source of the
interstellar gas? On what time scale(s) did it arrive?  Given this reservoir of cool gas, why does
the SFR remain low?

\section{The COS-Halos Survey: Motivations and Data}

Gas entering ETGs from their large-scale surroundings must pass through their circumgalactic medium
(CGM), where it is detectable with sensitive UV absorption-line tracers (limiting column density
$\NHI \simgt 10^{13} \cm{-2}$).  Quasar absorption line studies have attempted to link
\HI\ absorption (generally \Lya\ only) and nearby galaxies, but the results have been inconclusive,
suffering from small sample sizes and the inability to control for galaxy type and
SFR. \citet{chen-etal-01-Lya-imaging} showed that \Lya\ absorption is common around galaxies of a
wide range of morphological types, whereas \citet{chen-mulchaey-09-I-survey} showed that strong
\HI\ absorbers ($\NHI > 10^{14}\cmsq$) do not correlate with ETGs (see their
Figure~13). \citet{wakker-savage-09-OVI-HI-lowz} found ubiquitous \HI\ around $\gtrsim 0.1$\,L$^*$
galaxies at 350\kpc scales, but could not separate galaxies by type. Most recently,
\citet{prochaska-etal-11-OVI-HI} found that strong \Lya\ absorption ($W_r > 300\mA$) arises in
galaxy halos---including less massive ETGs---within $\rho \simeq 300\kpc$, but the majority of their
pairs lie at large radii (only 3 pairs at $\rho < 200\kpc$). Careful selection of QSO sightlines is
therefore crucial to study the dependence of CGM gas within 150\kpc, and spectroscopy of the host
galaxies is important to measure the current SFR and galaxy metallicities.

We designed the COS-Halos survey to probe the inner CGM of $L*$ galaxies at $z \sim 0.2$ and to test
the prevailing theoretical picture of ``hot'' and ``cold'' accretion, which might be related to the
observed star formation dichotomy in galaxies \citep{Keres:2005gba, db06}. COS-Halos has several
distinct advantages over previous work for assessing the amount and properties of CGM surrounding
ETGs: (1) it systematically covers impact parameters $0 - 150$ kpc and stellar masses $\log M_* = 10
- 11$ with galaxies from both color-magnitude sequences selected prior to knowledge of absorption,
(2) the wavelength range of COS and the galaxy redshifts ($z \gtrsim 0.1$) generally allow for
measurements of the weaker \HI\ Lyman series lines that give better column densities and line
broadening parameters than \lya\ alone, and (3) the galaxies are well characterized by ground-based
optical measurements of their stellar masses, SFRs, and metallicities. COS-Halos has previously
found that highly ionized \OVI\ in the halos of ETGs occurs at much lower incidence than in
star-forming galaxies, such that the presence of \OVI\ out to $\sim 150\kpc$ is related to star
formation \citep{tumlinson-etal-11-OVI-statistics}. We will present the full analysis of the
\HI\ absorption for our complete sample of 50 galaxies in a future paper (Tumlinson et al. 2012, in
preparation). In this Letter, we present results for \numetg\ ETGs from COS-Halos, showing that
their associated \HI\ gas is plentiful (\S~\ref{plentiful}), bound (\S~\ref{bound}), cold
(\S~\ref{cold}), and may represent a large mass of baryons in their CGM (\S~\ref{mass}).

The COS-Halos survey obtained 39 spectra of QSOs at $z < 1$ with the Cosmic Origins Spectrograph
\citep[COS;][]{green-etal-12-COS} onboard the Hubble Space Telescope (HST) under program GO-11598
(PI Tumlinson). These data were reduced, co-added, and continuum-normalized with procedures outlined
by \citet{Meiring2011}, \citet{tumlinson-etal-11-J1009-LLS}, and \citet{thom-etal-11-J0943-OVI}.  We
measured column densities for the absorption associated with targeted galaxies using both the apparent
optical depth technique \citep[AOD;][]{sembach-savage-92-EW} and iterative Voigt profile fitting
with the measured COS line spread function \citep{Ghavamian:09:1}. Our measurements for \HI\ column
density are weighted averages of unsaturated Lyman lines or profile fits to damped absorption. Where
all available Lyman series lines are saturated we set a lower limit to \NHI\ with the highest-order
detected Lyman line. The multi-component Voigt profile fitting yields $N$, $b$, and $v$ for
components within each system that are used in \S~\ref{cold}.

The medium-resolution Keck and Magellan spectroscopy used to derive the properties of the galaxies
in the sample (SFRs, metallicities, etc) are described fully by \citet{werk-etal-12-galaxies}. The
resulting sample contains 16 galaxies classified as passive and/or early type, based on their lack
of star-formation \citep[$sSFR = SFR / M_* < 10^{-11}$ M$_{\odot}$ yr$^{-1}$,
][]{werk-etal-12-galaxies}. We caution that these fields have not been uniformly surveyed for all
galaxies down to low luminosity limits to eliminate all other possible associations, but these
observations empirically describe the distribution of gas around ETGs as a function of impact
parameter, whatever its source.  Figure~\ref{fig1} shows an example of an ETG and star-forming
galaxy from our sample. The composite image from the Sloan Digital Sky Survey (SDSS) is shown, along
with the \Lya\ absorption owing to the galaxy. Both galaxies have similar \HI\ column densities and
absorption profiles.

\begin{figure}[!t]
\begin{center} 
\epsscale{1.2}
\plotone{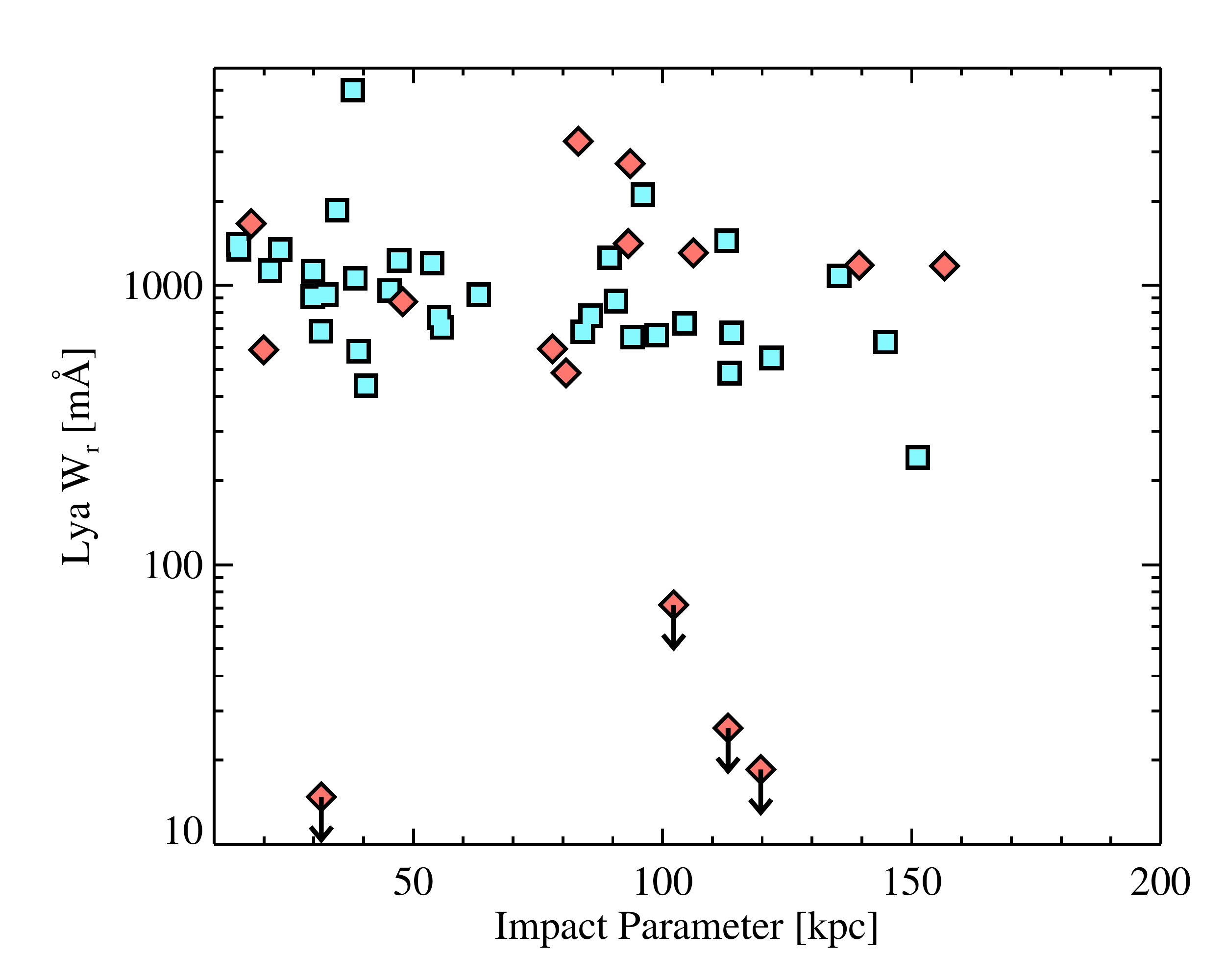}
\end{center} 
\caption{Ly$\alpha$ rest-frame equivalent width $W_r$ versus impact parameter $\rho$ for our
  COS-Halos sample of SF (blue) and ETG (red) galaxies. Note that only 15 ETGs appear in this
  figure, since in 1 case we do not have coverage of the \Lya\ transition.  Even with four low
  values seen in the ETGs, there is no statistically significant evidence of a dependence on galaxy
  type (red vs. blue).  \label{lya_wr_rho}}
\end{figure} 

\section{Strong \HI\ Is Common Around Early-type Galaxies}
\label{plentiful} 

Figure~\ref{lya_wr_rho} plots the \Lya\ rest-frame equivalent width $W_r$, as a function of impact
parameter $\rho$, and shows the remarkable similarity in \HI\ strength between the ETG (red) and SF
(blue) samples. A KS test on $W_r$ does not provide significant evidence against the null hypothesis
that the two distributions are drawn from the same parent population ($D = 0.32$, $P = 0.19$). The
four \Lya\ non-detections drive the difference in the KS statistic, and offer a hint that the
distributions may be slightly different, but we cannot separate the distributions in a statistical
fashion; a larger sample of ETGs will be needed to show if this difference is real.

We also calculated the ``covering fraction'' $f_{\rm hit}$ of \HI\ detections. Even if we
conservatively assume that the absorbers have the \NHI\ given by the lower limits for saturated
lines, we find that the covering fraction above $\logNHI = 16$ is $f_{\rm hit} = 0.21$ for SF
galaxies, and $f_{\rm hit} = 0.38$ for the ETGs. This column density has significance as the
threshold above which \cite{stewart11a} predicted essentially no absorbers surrounding halos at the
ETG mass scale, owing to the transition between hot and cold accretion. Our covering fraction result
rules out $f_{\rm hit} = 0.01$ at $>$99.9\% confidence. The corresponding covering fractions at
$\logNHI > 15$ are $f_{\rm hit} = 0.64$ (SF) and $f_{\rm hit} = 0.56$ (ETGs), again consistent to
within the statistical errors of 0.15-0.2. Thus we find that strong \HI\ absorption is common around
the ETGs, and that to the limits of our survey, it does not appear significantly less often or
weaker than in the halos of star-forming galaxies.


\section{The Strong \HI\ is Bound to Its Host Galaxy} 
\label{bound} 

\begin{figure}[!t]
\begin{center} 
\epsscale{1.2}
\plotone{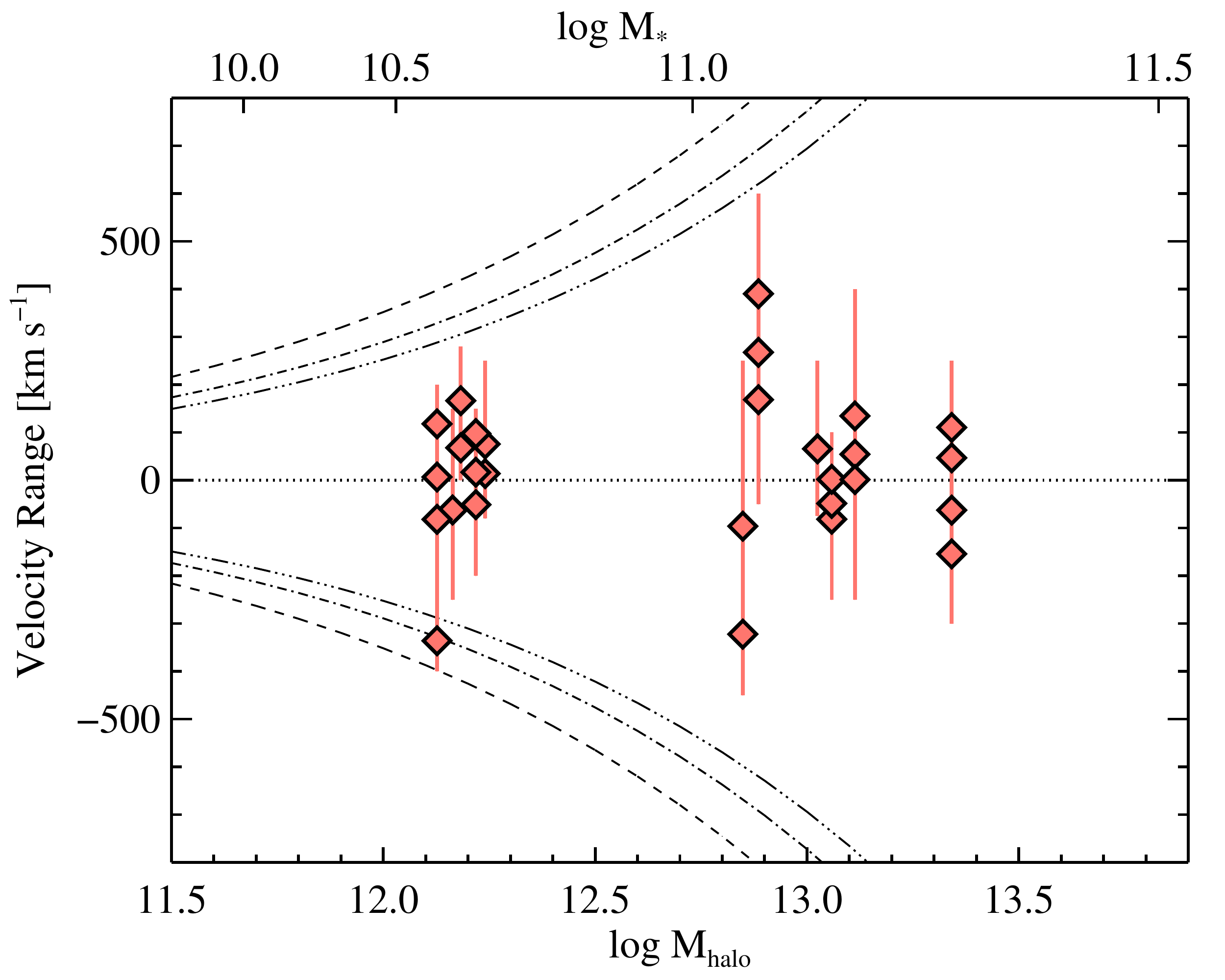}
\end{center} 
\caption{The full velocity range occupied by the detected HI absorption, with respect to the galaxy
  systemic redshift ($v = 0$), plotted versus the inferred dark-matter halo mass for our
  sample. The fitted component positions are indicated by the diamonds, while the full velocity
  ranges are shown by the vertical red bar. The three dashed lines mark the halo escape velocity at
  50, 100, and 150 kpc, from outside to inside. All the \HI\ detected in the ETGs is well inside the
  halo escape velocity, suggesting that is bound to the host galaxies. \label{lya_kinematics}}
\end{figure}

The detected \HI\ shows a strong relationship to the targeted ETGs. It is found at separations
consistent with being well inside the virial radius ($R_{\rm vir} \sim 300$ kpc), and it is closely
associated with the stars in velocity space. Figure~\ref{lya_kinematics} shows the {\it full}
velocity range of the detected \HI\ absorption with respect to the systemic velocity of the galaxy's
stars (set to $v = 0$). The relative errors between absorption and galaxy redshift (i.e. between the
wavelength solutions of the optical and COS spectroscopy) is $\sim 30\kms$. The filled symbols mark
centroid velocities of the fitted components, while the range bars show the full width at zero
absorption, a measure of the greatest kinematic extent of the detected \HI. These velocities are
plotted with respect to the inferred dark matter halo mass, which is derived from the
photometrically-estimated stellar mass using the mean relation from \cite{Moster2010}. The dashed
curves show the escape velocity from the halos as a function of mass, at distances of 50, 100, and
150 kpc (from outside to inside). Nearly all the detected \HI\ is found at velocities below the escape
velocity. This is true even if we increase the velocity ranges by $\sqrt 3$ to account for
unconstrained projection effects (since relative Doppler shifts are a sightline-projected lower
limit to the true 3D space motion). It is also possible that some of the detected ionized gas is in
fact bound to satellite galaxies in the vicinity of the targeted galaxy\footnote{This is very likely
  in the two systems with $\log$ \NHI $>10^{19}$.}; if so, the kinematics suggest that the
satellites are themselves bound and would be counted among the CGM mass budget for the host galaxy.
We conclude that the detected HI is predominantly bound to its host galaxy, and not escaping at the
time it is observed.

\section{The Strong \HI\ Is Cold Compared with Halo Virial Temperatures} 
\label{cold} 

\begin{figure}[!t]
\begin{center} 
\epsscale{1.2}
\plotone{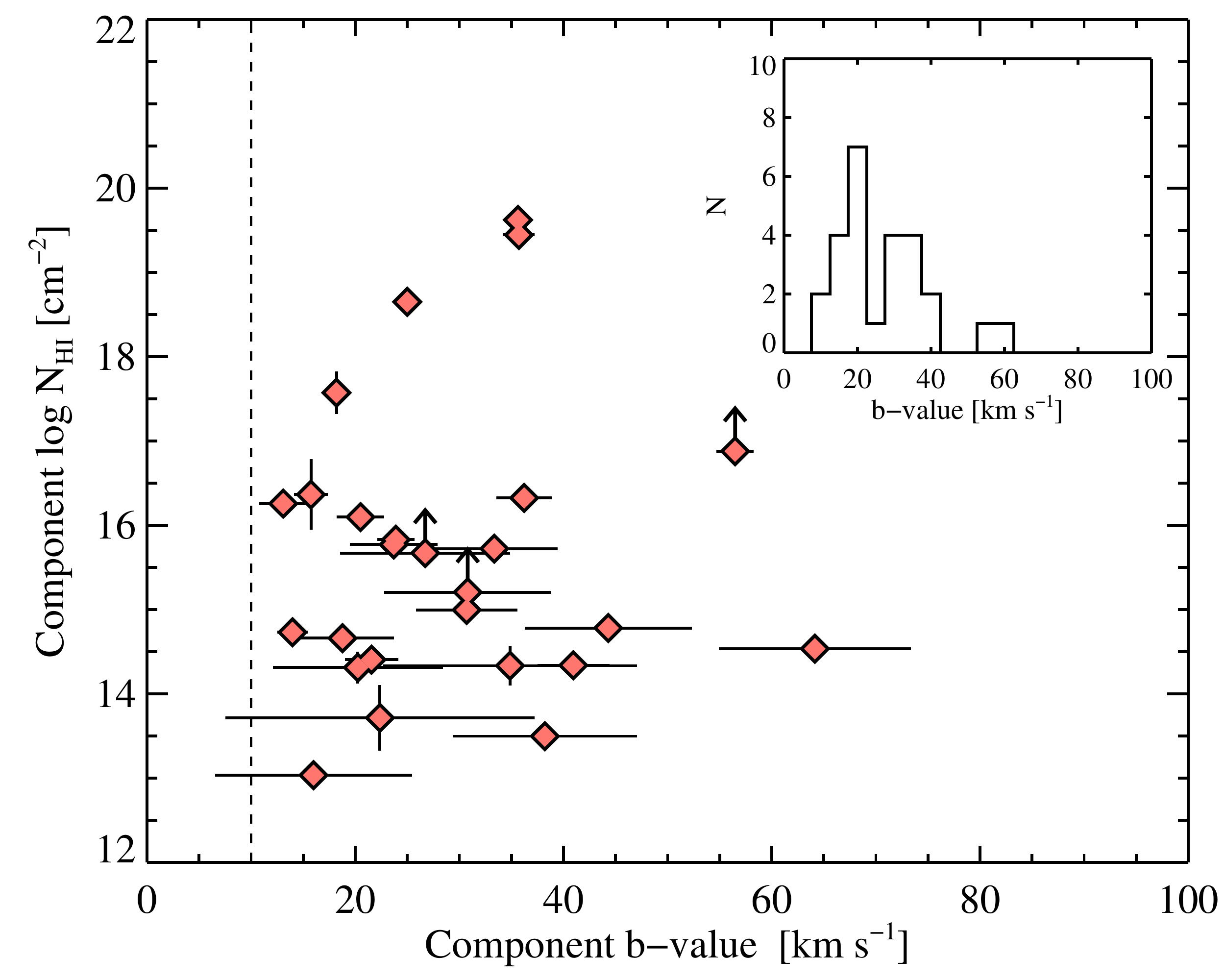}
\end{center} 
\caption{On a component-by-component basis, the \NHI\ and Doppler $b$ parameter 
for \HI\ around ETGs. Even if we assume that there are no non-thermal contributions to the 
line-broadening, most of the detected \HI\ has implied temperatures of $T \lesssim 2 \times 10^5$ K. 
\label{nhi_vs_b}}
\end{figure}

Because COS-Halos typically covers multiple Lyman series lines at $z \gtrsim 0.2$
(Figure~\ref{fig1}), we can often derive line-broadening parameters that are not available when
relying solely on \lya\ (which is almost always saturated). We perform profile fits to derive
\NHI\ and Doppler $b$ for each component, where possible. These line widths provide robust upper
limits on the gas temperature under the assumption that the broadening is purely thermal, $b \equiv
\sqrt{2kT/m_H} $. From the measured Doppler parameters (Figure~\ref{nhi_vs_b}) we can firmly
establish temperature upper limits of $T \lesssim 2 \times 10^5\K$ ($b_{\rm therm} < 60$ \kms), with
80\% at $T < 10^5$ K ($b_{\rm therm} = 40$ \kms).  These line widths are upper limits based on the
observed profiles, and can only decrease if the observed profile turns out to be composed of
narrower unresolved components, or is significantly affected by non-thermal broadening \citep[which
  has shown to be common;][]{thom-chen-08-II-OVI-absorbers,tripp-etal-08-OVI}\footnote{There may
  also be shallow ``broad Lyman $\alpha$'' (BLA) absorption hidden in the profiles; we cannot draw
  any conclusions about the presence or absence an additional hot gas component in ETGs.}.
  
We therefore conclude that the detected \HI\ traces mainly ``cool'' (possibly photoionized) gas, at
temperatures well below the $\gtrsim 10^6\K$ virial temperatures of halos in this mass range. These
temperatures resemble the expectations for the ``cold mode'' of galaxy gas accretion. Together with
the finding that the detected gas is plentiful and bound, this finding suggests that the deposition
of cold gas into galaxy halos (from the inside or the outside) and/or the formation of clouds at
$\ll T_{\rm vir}$ are not fully quenched even in galaxies that have ceased to form stars.

\section{Implications for the CGM Mass}
\label{mass} 

The total mass of cool gas in ETG halos is of interest in comparison to their stellar masses and
interstellar gas reservoirs. Having measured the covering fraction and typical column density, we
can estimate the detected \HI\ mass from physical arguments and simple models. The simplest possible
mass estimate comes from multiplying a surface density by an area:
\begin{eqnarray*}
M_{\rm HI}  & = &\pi R^2 \langle N_{\rm HI} \rangle m_{\rm H}  f_{\rm hit} \\
           & \geq & 2.8 \times 10^6  \left({ N_{\rm HI} \over  10^{16}}\right)  \left({ f_{\rm hit} \over
               0.5}\right) \left({ R } \over {150\kpc} \right)^2 M_\odot
  \label{masseq}
\end{eqnarray*}
where we have taken a typical $N_{\rm HI} = 10^{16}\, {\rm cm}^{-2}$, at which the hit rate is
$f_{\rm hit} = 0.5$ for ETGs, and $\rho =150\kpc$. This mass of \HI\ is strictly a lower limit
because we have taken the typical minimum \NHI\ permitted by the profiles of saturated lines and
surveyed a radius that may not encompass the full CGM mass
\citep[c.f. ][]{prochaska-etal-11-OVI-HI}. The corresponding total mass of CGM hydrogen is $M_{\rm H}
= M_{\rm HI} / f_{\rm HI}$, where $f_{\rm HI}$ is the typical neutral fraction in the gas. Thus the
ionization correction is the major factor in setting the total mass, but the ionization conditions,
and thus the total mass of the CGM gas, depend on the unknown temperature and density of the
detected material. Modestly overdense gas in the CGM should be exposed to the cosmic ionizing
background radiation, which will further suppress the neutral faction and imply a much larger mass.

\begin{figure}[!t]
\begin{center} 
\epsscale{1.2}
\plotone{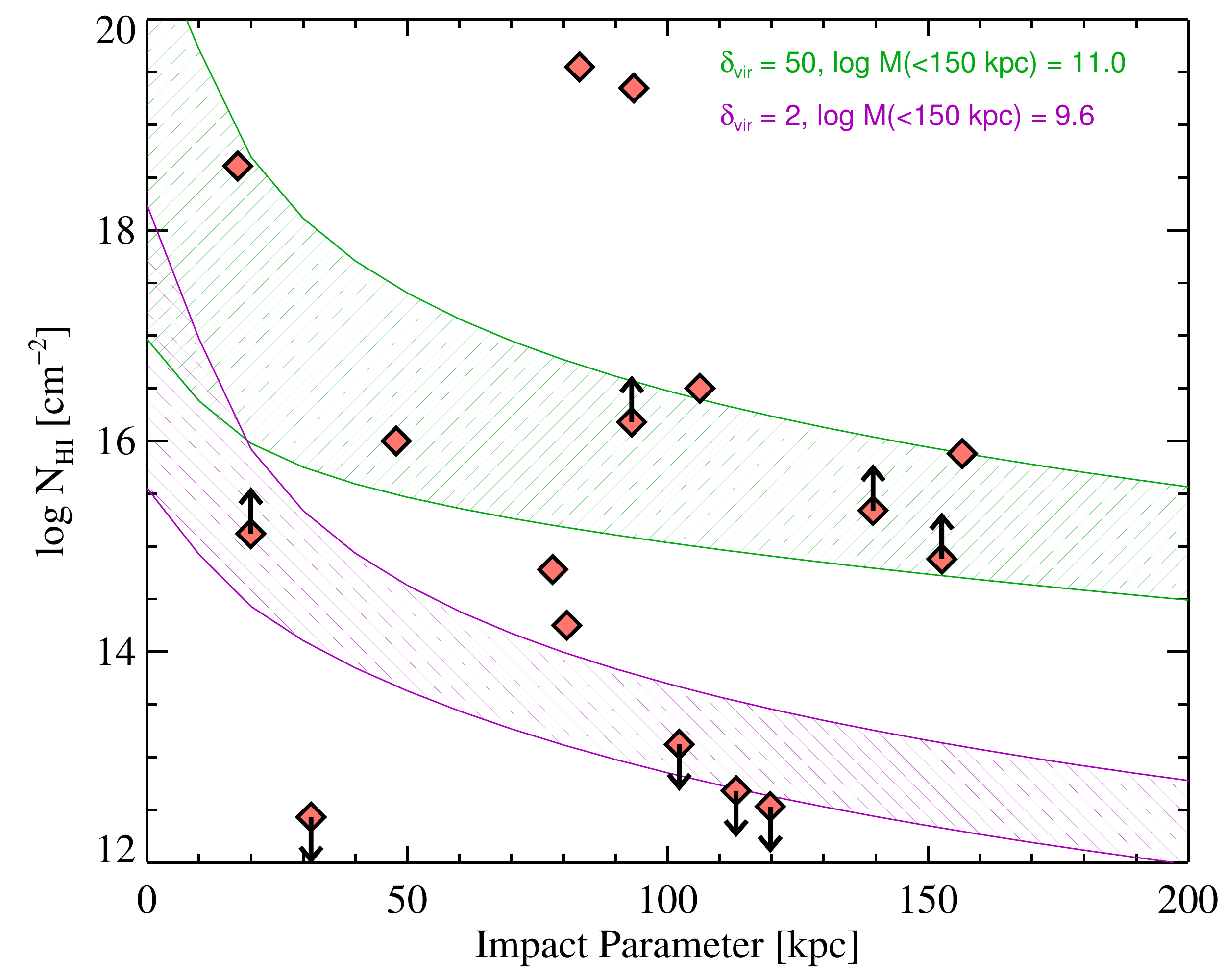}
\end{center} 
\caption{\NHI\ versus impact parameter $\rho$ for all \numetg\ ETGs. Lower limits are systems for
  which all available Lyman lines are saturated.  The model profiles show two simple halo models for
  overdensity normalization $\delta _{\rm vir} = 50$ (upper pair, green) and $\delta _{\rm vir} = 2$
  (lower pair, purple).  Within each set of curves the lower curve corresponds to $T = 10^5$ K,
  while the upper bounding curves are for $T = 10^{4.3}$ K; these are the two relevant temperature
  limits from the kinematics analysis above. The total gas masses inside 150 kpc are $\log M$($<$150
  kpc)$ = 9.6 - 11$. \label{lya_nhi_rho}}
\end{figure} 

To estimate what CGM masses are possible for the ETGs we show a simple halo model in
Figure~\ref{lya_nhi_rho} that uses a power-law gas density profile exposed to a uniform ionizing
background and held at fixed temperature. The gas density follows a power-law dependence with
radius, $n \propto (R / R_{\rm vir})^{-\alpha}$, with $\alpha = 2$, normalized to a specific cosmic
overdensity $\delta _0$ at $R_{vir}$. The photoionization equilibrium solutions were derived using
Cloudy \citep{Ferland1998} models and including ionization by the Haardt \& Madau photoionizing
background (for $z = 0.2$) and the temperature-dependent collisional ionization equilibrium
(CIE). We neglect any photoionizing contribution from the nearby ETGs since their star-formation
rates are not high enough to contribute significantly to the 1~Ryd background in their halos
\citep{tumlinson-etal-11-J1009-LLS}. Overdensities of $\delta _0 = 20 - 50$ are required to match
the ETG \HI\ detections with $\logNHI \sim 16$. This profile gives a total mass of $\log M_{\rm
  H} \sim 11$. The lower model adopts $\delta _0 = 2$, and is a better match to the
non-detections. Note that the results depend less on the chosen temperature than on the density
normalization. From these simple models we conclude that the detected CGM is consistent with a mass
$M_{CGM} \gtrsim 10^{9-11}$ $M_{\odot}$. This mass is significant by comparison with the {\em
  interstellar} gas masses detected in other ETGs (\S~1). At the high end these masses are
potentially significant reservoirs of ``missing baryons'' in the budgets of these massive halos. We
emphasize that these simple models are not rigorous derivations of the ionization correction, from
which the actual gas column densities $N_H$ can be inferred. We will publish separately a full
analysis of the gas column densities based on models calibrated to the many available metal lines in
these absorbers (J.~K. Werk et al. in preparation). Rather, these simple estimates show that ETGs
harbor a significant mass of cold, bound gas in their CGM.

\section{Implications} 

The COS-Halos survey has detected significant quantities of cool, likely photoionized gas bound to
the halos of massive early-type galaxies. The cool gas does not appear to be correlated with
star-formation, in stark contrast with \OVI, which is strongly dependent on the presence or absence
of star-formation \citep{tumlinson-etal-11-OVI-statistics}. The cool gas reservoir may contribute
$\gtrsim 10^{9-11} M_{\odot}$ of gas to the halos of ETGs, comparable to the interstellar reservoirs
(\S~1), and exists at well below the halo virial temperatures. This material is typically well
within the escape velocities of the host galaxies, so it may be re-accreted by the galaxies and
eventually used for star-formation. The question then becomes, why does the gas not do so?

These ETGs have been quenched, in the usual sense that they are no longer forming stars. These
strong \HI\ absorbers and the cool CGM they trace were predicted to disappear at the end of cold
accretion, which should not operate in halos at this mass \citep{stewart11a}.  If $M_{\rm CGM} \sim
10^{10} M_{\odot}$ is typical of the star forming galaxies in COS-Halos as well as the ETGs, then
the ETGs probably have less mass in their CGM {\em relative to their stellar and halo masses} than
the SF galaxies (at least in cool gas). This might be the signature of quenching that can suppress
but not completely remove cold gas from the halo, or that the gas can reaccrete long after
quenching. It is also possible that the quenching mechanism {\it did} remove all the CGM gas, after
which cold clouds re-accrete from the IGM or reform from thermal instabilities in the hot corona.

Models that attempt to explain the formation and evolution of ETGs will need to account for this
significant budget of cold gas in their halos. While it is possible that some of this gas is the
direct product of ejection by supernovae while star-formation was occurring, it is perhaps more
likely to be gas cooled from their hot halos or falling in from the IGM. In this case, preventive
feedback may act to slow its accretion, allowing only small budgets of interstellar gas to enter the
ISM and form stars. Some of it (especially the stronger systems) may be bound to satellites that are
themselves bound to the host, and thus present or future contributors to the diffuse CGM. In any
case, our findings indicate that the quiescence of star formation in ETGs generally cannot be
attributed exclusively to the quenching of gas in their circumgalactic medium.

\acknowledgments

Support for program GO11598 was provided by NASA through a grant from the Space Telescope Science
Institute, which is operated by the Association of Universities for Research in Astronomy, Inc.,
under NASA contract NAS 5-26555. Some of the data presented herein were obtained at the W.M. Keck
Observatory, which is operated as a scientific partnership among the California Institute of
Technology, the University of California and the National Aeronautics and Space Administration. TMT
appreciates support from NASA grant NNX08AJ44G. MSP acknowledges support from the Southern
California Center for Galaxy Evolution, a multi-campus research program funded by the University of
California Office of Research.



\begin{thebibliography}{36}
\expandafter\ifx\csname natexlab\endcsname\relax\def\natexlab#1{#1}\fi

\bibitem[{{Baldry} {et~al.}(2004){Baldry}, {Glazebrook}, {Brinkmann},
  {Ivezi{\'c}}, {Lupton}, {Nichol}, \& {Szalay}}]{baldry04}
{Baldry}, I.~K., {Glazebrook}, K., {Brinkmann}, J., {et~al.} 2004, \apj, 600,
  681

\bibitem[{{Bell} {et~al.}(2004){Bell}, {Wolf}, {Meisenheimer}, {Rix}, {Borch},
  {Dye}, {Kleinheinrich}, {Wisotzki}, \& {McIntosh}}]{bell04}
{Bell}, E.~F., {Wolf}, C., {Meisenheimer}, K., {et~al.} 2004, \apj, 608, 752

\bibitem[{{Bland-Hawthorn} {et~al.}(2007){Bland-Hawthorn}, {Sutherland},
  {Agertz}, \& {Moore}}]{bland-hawthorn-etal-07-MS-Halpha}
{Bland-Hawthorn}, J., {Sutherland}, R., {Agertz}, O., \& {Moore}, B. 2007,
  \apjl, 670, L109

\bibitem[{{Chen} \& {Mulchaey}(2009)}]{chen-mulchaey-09-I-survey}
{Chen}, H., \& {Mulchaey}, J.~S. 2009, \apj, 701, 1219

\bibitem[{{Chen} {et~al.}(2001){Chen}, {Lanzetta}, {Webb}, \&
  {Barcons}}]{chen-etal-01-Lya-imaging}
{Chen}, H.-W., {Lanzetta}, K.~M., {Webb}, J.~K., \& {Barcons}, X. 2001, \apj,
  559, 654

\bibitem[{{Crocker} {et~al.}(2011){Crocker}, {Bureau}, {Young}, \&
  {Combes}}]{crocker11}
{Crocker}, A.~F., {Bureau}, M., {Young}, L.~M., \& {Combes}, F. 2011, \mnras,
  410, 1197

\bibitem[{{Davis} {et~al.}(2011){Davis}, {Alatalo}, {Sarzi}, {Bureau}, {Young},
  {Blitz}, {Serra}, {Crocker}, {Krajnovi{\'c}}, {McDermid}, {Bois}, {Bournaud},
  {Cappellari}, {Davies}, {Duc}, {de Zeeuw}, {Emsellem}, {Khochfar},
  {Kuntschner}, {Lablanche}, {Morganti}, {Naab}, {Oosterloo}, {Scott}, \&
  {Weijmans}}]{davis11}
{Davis}, T.~A., {Alatalo}, K., {Sarzi}, M., {et~al.} 2011, \mnras, 417, 882

\bibitem[{{Dekel} \& {Birnboim}(2006)}]{db06}
{Dekel}, A., \& {Birnboim}, Y. 2006, \mnras, 368, 2

\bibitem[{{Faber} {et~al.}(2007){Faber}, {Willmer}, {Wolf}, {Koo}, {Weiner},
  {Newman}, {Im}, {Coil}, {Conroy}, {Cooper}, {Davis}, {Finkbeiner}, {Gerke},
  {Gebhardt}, {Groth}, {Guhathakurta}, {Harker}, {Kaiser}, {Kassin},
  {Kleinheinrich}, {Konidaris}, {Kron}, {Lin}, {Luppino}, {Madgwick},
  {Meisenheimer}, {Noeske}, {Phillips}, {Sarajedini}, {Schiavon}, {Simard},
  {Szalay}, {Vogt}, \& {Yan}}]{faber07}
{Faber}, S.~M., {Willmer}, C.~N.~A., {Wolf}, C., {et~al.} 2007, \apj, 665, 265

\bibitem[{Ferland {et~al.}(1998)Ferland, Korista, Verner, Ferguson, Kingdon, \&
  Verner}]{Ferland1998}
Ferland, G.~J., Korista, K.~T., Verner, D.~A., {et~al.} 1998, The Publications
  of the Astronomical Society of the Pacific, 110, 761

\bibitem[{{Ghavamian} {et~al.}(2009){Ghavamian}, {Aloisi}, {Lennon}, {Hartig},
  {Kriss}, {Oliveira}, {Massa}, {Keyes}, {Proffitt}, {Delker}, \&
  {Osterman}}]{Ghavamian:09:1}
{Ghavamian}, P., {Aloisi}, A., {Lennon}, D., {et~al.} 2009, {Preliminary
  Characterization of the Post- Launch Line Spread Function of COS}, Tech. rep.

\bibitem[{{Gonz{\'a}lez}(1993)}]{gonzalez93}
{Gonz{\'a}lez}, J.~J. 1993, PhD thesis, Thesis (PH.D.)--UNIVERSITY OF
  CALIFORNIA, SANTA CRUZ, 1993.Source: Dissertation Abstracts International,
  Volume: 54-05, Section: B, page: 2551.

\bibitem[{{Green} {et~al.}(2012){Green}, {Froning}, {Osterman}, {Ebbets},
  {Heap}, {Leitherer}, {Linsky}, {Savage}, {Sembach}, {Shull}, {Siegmund},
  {Snow}, {Spencer}, {Stern}, {Stocke}, {Welsh}, {B{\'e}land}, {Burgh},
  {Danforth}, {France}, {Keeney}, {McPhate}, {Penton}, {Andrews},
  {Brownsberger}, {Morse}, \& {Wilkinson}}]{green-etal-12-COS}
{Green}, J.~C., {Froning}, C.~S., {Osterman}, S., {et~al.} 2012, \apj, 744, 60

\bibitem[{{Kaviraj}(2010)}]{kaviraj-10-ETG-SF}
{Kaviraj}, S. 2010, \mnras, 408, 170

\bibitem[{Kere{\v s} {et~al.}(2005)Kere{\v s}, Katz, Weinberg, \&
  Dav{\'e}}]{Keres:2005gba}
Kere{\v s}, D., Katz, N., Weinberg, D.~H., \& Dav{\'e}, R. 2005, Monthly
  Notices of the Royal Astronomical Society, 363, 2

\bibitem[{{Lees} {et~al.}(1991){Lees}, {Knapp}, {Rupen}, \&
  {Phillips}}]{lees91}
{Lees}, J.~F., {Knapp}, G.~R., {Rupen}, M.~P., \& {Phillips}, T.~G. 1991, \apj,
  379, 177

\bibitem[{{Meiring} {et~al.}(2011){Meiring}, {Tripp}, {Prochaska}, {Tumlinson},
  {Werk}, {Jenkins}, {Thom}, {O\'Meara}, \& {Sembach}}]{Meiring2011}
{Meiring}, J.~D., {Tripp}, T.~M., {Prochaska}, J.~X., {et~al.} 2011, \apj, 732,
  35

\bibitem[{{Moster} {et~al.}(2010){Moster}, {Somerville}, {Maulbetsch}, {van den
  Bosch}, {Macci{\`o}}, {Naab}, \& {Oser}}]{Moster2010}
{Moster}, B.~P., {Somerville}, R.~S., {Maulbetsch}, C., {et~al.} 2010, \apj,
  710, 903

\bibitem[{{Oosterloo} {et~al.}(2010){Oosterloo}, {Morganti}, {Crocker},
  {J{\"u}tte}, {Cappellari}, {de Zeeuw}, {Krajnovi{\'c}}, {McDermid},
  {Kuntschner}, {Sarzi}, \& {Weijmans}}]{oosterloo10}
{Oosterloo}, T., {Morganti}, R., {Crocker}, A., {et~al.} 2010, \mnras, 409, 500

\bibitem[{{Prochaska} {et~al.}(2011){Prochaska}, {Weiner}, {Chen}, {Mulchaey},
  \& {Cooksey}}]{prochaska-etal-11-OVI-HI}
{Prochaska}, J.~X., {Weiner}, B., {Chen}, H.-W., {Mulchaey}, J., \& {Cooksey},
  K. 2011, \apj, 740, 91

\bibitem[{{Putman} {et~al.}(2011){Putman}, {Saul}, \&
  {Mets}}]{putman-etal-11-head-tail-CHVCs}
{Putman}, M.~E., {Saul}, D.~R., \& {Mets}, E. 2011, \mnras, 418, 1575

\bibitem[{{Sembach} \& {Savage}(1992)}]{sembach-savage-92-EW}
{Sembach}, K.~R., \& {Savage}, B.~D. 1992, \apjs, 83, 147

\bibitem[{{Serra} {et~al.}(2011){Serra}, {Oosterloo}, {Morganti}, {Alatalo},
  {Blitz}, {Bois}, {Bournaud}, {Bureau}, {Cappellari}, {Crocker}, {Davies},
  {Davis}, {de Zeeuw}, {Duc}, {Emsellem}, {Khochfar}, {Krajnovic},
  {Kuntschner}, {Lablanche}, {McDermid}, {Naab}, {Sarzi}, {Scott}, {Trager},
  {Weijmans}, \& {Young}}]{serra11}
{Serra}, P., {Oosterloo}, T., {Morganti}, R., {et~al.} 2011, ArXiv e-prints

\bibitem[{{Springel} {et~al.}(2005){Springel}, {Di Matteo}, \&
  {Hernquist}}]{sdh05}
{Springel}, V., {Di Matteo}, T., \& {Hernquist}, L. 2005, \apjl, 620, L79

\bibitem[{{Stewart} {et~al.}(2011){Stewart}, {Kaufmann}, {Bullock}, {Barton},
  {Maller}, {Diemand}, \& {Wadsley}}]{stewart11a}
{Stewart}, K.~R., {Kaufmann}, T., {Bullock}, J.~S., {et~al.} 2011, \apjl, 735,
  L1

\bibitem[{{Thom} \& {Chen}(2008)}]{thom-chen-08-II-OVI-absorbers}
{Thom}, C., \& {Chen}, H.-W. 2008, \apjs, 179, 37

\bibitem[{{Thom} {et~al.}(2011){Thom}, {Werk}, {Tumlinson}, {Prochaska},
  {Meiring}, {Tripp}, \& {Sembach}}]{thom-etal-11-J0943-OVI}
{Thom}, C., {Werk}, J.~K., {Tumlinson}, J., {et~al.} 2011, \apj, 736, 1

\bibitem[{{Thomas} {et~al.}(2010){Thomas}, {Maraston}, {Schawinski}, {Sarzi},
  \& {Silk}}]{thomas10}
{Thomas}, D., {Maraston}, C., {Schawinski}, K., {Sarzi}, M., \& {Silk}, J.
  2010, \mnras, 404, 1775

\bibitem[{{Tonnesen} \& {Bryan}(2009)}]{tb09}
{Tonnesen}, S., \& {Bryan}, G.~L. 2009, \apj, 694, 789

\bibitem[{{Trager} {et~al.}(2000){Trager}, {Faber}, {Worthey}, \&
  {Gonz{\'a}lez}}]{tfw+00}
{Trager}, S.~C., {Faber}, S.~M., {Worthey}, G., \& {Gonz{\'a}lez}, J.~J. 2000,
  \aj, 119, 1645

\bibitem[{{Tripp} {et~al.}(2008){Tripp}, {Sembach}, {Bowen}, {Savage},
  {Jenkins}, {Lehner}, \& {Richter}}]{tripp-etal-08-OVI}
{Tripp}, T.~M., {Sembach}, K.~R., {Bowen}, D.~V., {et~al.} 2008, \apjs, 177, 39

\bibitem[{{Tumlinson} {et~al.}(2011{\natexlab{a}}){Tumlinson}, {Werk}, {Thom},
  {Meiring}, {Prochaska}, {Tripp}, {O'Meara}, {Okrochkov}, \&
  {Sembach}}]{tumlinson-etal-11-J1009-LLS}
{Tumlinson}, J., {Werk}, J.~K., {Thom}, C., {et~al.} 2011{\natexlab{a}}, \apj,
  733, 111

\bibitem[{{Tumlinson} {et~al.}(2011{\natexlab{b}}){Tumlinson}, {Thom}, {Werk},
  {Prochaska}, {Tripp}, {Weinberg}, {Peeples}, {O\'Meara}, {Oppenheimer},
  {Meiring}, {Katz}, {Dav{\'e}}, {Ford}, \&
  {Sembach}}]{tumlinson-etal-11-OVI-statistics}
{Tumlinson}, J., {Thom}, C., {Werk}, J.~K., {et~al.} 2011{\natexlab{b}},
  Science, 334, 948

\bibitem[{{Wakker} \& {Savage}(2009)}]{wakker-savage-09-OVI-HI-lowz}
{Wakker}, B.~P., \& {Savage}, B.~D. 2009, \apjs, 182, 378

\bibitem[{{Werk} {et~al.}(2012){Werk}, {Prochaska}, {Thom}, {Tumlinson},
  {Tripp}, {O'Meara}, \& {Meiring}}]{werk-etal-12-galaxies}
{Werk}, J.~K., {Prochaska}, J.~X., {Thom}, C., {et~al.} 2012, \apjs, 198, 3

\bibitem[{{Young} {et~al.}(2011){Young}, {Bureau}, {Davis}, {Combes},
  {McDermid}, {Alatalo}, {Blitz}, {Bois}, {Bournaud}, {Cappellari}, {Davies},
  {de Zeeuw}, {Emsellem}, {Khochfar}, {Krajnovi{\'c}}, {Kuntschner},
  {Lablanche}, {Morganti}, {Naab}, {Oosterloo}, {Sarzi}, {Scott}, {Serra}, \&
  {Weijmans}}]{young11}
{Young}, L.~M., {Bureau}, M., {Davis}, T.~A., {et~al.} 2011, \mnras, 414, 940

\end{thebibliography}
\bibliographystyle{apj}

\end{document}